# Correction of inertial navigation system's errors by the help of video-based navigator based on Digital Terrarium Map.

**Kupervasser O. Yu.[1], Rubinstein A.A.**
Transist Video LLC, Skolkovo resident
[1]olegkup@yahoo.com

## Abstract

This paper deals with the error analysis of a novel navigation algorithm that uses as input the sequence of images acquired from a moving camera and a Digital Terrain (or Elevation) Map (DTM/DEM). More specifically, it has been shown that the optical flow derived from two consecutive camera frames can be used in combination with a DTM to estimate the position, orientation and ego-motion parameters of the moving camera. As opposed to previous works, the proposed approach does not require an intermediate explicit reconstruction of the 3D world. In the present work the sensitivity of the algorithm outlined above is studied. The main sources for errors are identified to be the optical-flow evaluation and computation, the quality of the information about the terrain, the structure of the observed terrain and the trajectory of the camera. By assuming appropriate characterization of these error sources, a closed form expression for the uncertainty of the pose and motion of the camera is first developed and then the influence of these factors is confirmed using extensive numerical simulations. The main conclusion of this paper is to establish that the proposed navigation algorithm generates accurate estimates for reasonable scenarios and error sources, and thus can be effectively used as part of a navigation system of autonomous vehicles.

## 1  Introduction

Vision-based algorithms has been a major research issue during the past decades. Two common approaches for the navigation problem are: *landmarks* and *ego-motion integration*. In the landmarks approach several features are located on the image-plane and matched to their known 3D location. Using the 2D and 3D data the camera's pose can be derived. Few examples for such algorithms are [2], [3]. Once the landmarks were found, the pose derivation is simple and can achieve quite accurate estimates. The main difficulty is the detection of the features and their correct matching to the landmarks set.

In ego-motion integration approach the motion of the camera with respect to itself is estimated. The ego-motion can be derived from the optical-flow field, or from instruments such as accelerometers and gyroscopes. Once the ego-motion was obtained, one can integrate this motion to derive the camera's path. One of the factors that make this approach attractive is that no specific features need to be detected, unlike the previous approach. Several ego-motion estimation algorithms can be found in [4], [5], [6], [7]. The weakness of ego-motion integration comes from the fact that small errors are accumulated during the integration process. Hence, the estimated camera's path is drifted and the pose estimation accuracy decrease along time. If such approach is used it would be desirable to reduce the drift by activating, once in a while, an additional algorithm that estimates the pose directly. In [8], such navigation-system is being suggested. In that work, like in this work, the drift is being corrected using a Digital Terrain Map (DTM). The DTM is a discrete representation of the observed ground's topography. It contains the altitude over the sea level of the terrain for each geographical location. In [8] a patch from the ground was reconstructed using

`structure-from-motion' (SFM) algorithm and was matched to the DTM in order to derive the camera's pose. Using SFM algorithm which does not make any use of the information obtained from the DTM but rather bases its estimate on the flow-field alone, positions their technique under the same critique that applies for SFM algorithms [1].

The algorithm presented in this work does not require an intermediate explicit reconstruction of the 3D world. By combining the DTM information directly with the images information it is claimed that the algorithm is well-conditioned and generates accurate estimates for reasonable scenarios and error sources. In the present work this claim is explored by performing an error analysis on the algorithm outlined above. By assuming appropriate characterization of these error sources, a closed form expression for the uncertainty of the pose and motion of the camera is first developed and then the influence of different factors is studied using extensive numerical simulations.

Comparison of the corrected position of the object, measured by an independent navigation system DGPS, with the calculated position of the object would estimate the real effectiveness of navigation corrections. The correspondent investigation for described method was made during flight in Galilee in Israel [9]. The position error was about 25 meter and angle error was about 1.5 degree.

## 2  Problem Definition and Notations

The problem can be briefly described as follows: At any given time instance $t$, a coordinates system $C(t)$ is fixed to a camera in such a way that the $Z$-axis coincides with the optical-axis and the origin coincides with the camera's projection center. At that time instance the camera is located at some geographical location $p(t)$ and has a given orientation $R(t)$ with respect to a global coordinates system $W$ ($p(t)$ is a 3D vector, $R(t)$ is an orthonormal rotation matrix). $p(t)$ and $R(t)$ define the transformation from the camera's frame $C(t)$ to the world's frame $W$, where if $^C v$ and $^W v$ are vectors in $C(t)$ and $W$ respectively, then $^W v = R(t)^C v + p(t)$.

Consider now two sequential time instances $t_1$ and $t_2$: the transformation from $C(t_1)$ to $C(t_2)$ is given by the translation vector $\Delta p(t_1,t_2)$ and the rotation matrix $\Delta R(t_1,t_2)$, such that $^{C(t_2)} v = \Delta R(t_1,t_2)^{C(t_1)} v + \Delta p(t_1,t_2)$. A rough estimate of the camera's pose at $t_1$ and of the ego-motion between the two time instances - $p_E(t_1)$, $R_E(t_1)$, $\Delta p_E(t_1,t_2)$ and $\Delta R_E(t_1,t_2)$ - are supplied (the subscript letter ``$E$'' denotes that this is an estimated quantity).

Also supplied is the optical-flow field: $\{u_i(t_k)\}$ ($i=1...n$, $k=1,2$). For the $i$'th feature, $u_i(t_1) \in \mathbb{R}^2$ and $u_i(t_2) \in \mathbb{R}^2$ represent its locations at the first and second frame respectively.

Using the above notations, the objective of the proposed algorithm is to estimate the true camera's pose and ego-motion: $p(t_1)$, $R(t_1)$, $\Delta p(t_1,t_2)$ and $\Delta R(t_1,t_2)$, using the optical-flow field $\{u_i(t_k)\}$, the DTM and the initial-guess: $p_E(t_1)$, $R_E(t_1)$, $\Delta p_E(t_1,t_2)$ and $\Delta R_E(t_1,t_2)$.

## 3.  The Navigation Algorithm

The following section describes a navigation algorithm which estimate the above mentioned parameters. The pose and ego-motion of the camera are derived using a DTM and the optical-flow field of two consecutive frames. Unlike the landmarks approach no specific features should be detected and matched. Only the correspondence between the two consecutive images should be found in order to derive the optical-flow field. As was mentioned in the previous section, a rough estimate of the required parameters is supplied as an input. Nevertheless, since the algorithm only use this input as an initial guess and re-calculate the pose and ego-motion directly, no integration of previous errors will take place and accuracy will be preserved.

The new approach is founded on the following observation. Since the DTM supplies information about the structure of the observed terrain, depth of observed features is being dictated by the camera's pose. Hence, given the pose and ego-motion of the camera, the optical-flow field can be uniquely determined. The objective of the algorithm will be finding the pose and ego-motion which lead to an optical-flow field as close as possible to the given flow field.

A single vector from the optical-flow field will be used to define a constraint for the camera's pose and ego-motion. Let $^WG \in \mathbb{R}^3$ be a location of a ground feature point in the 3D world. At two different time instances $t_1$ and $t_2$, this feature point is projected on the image-plane of the camera to the points $u(t_1)$ and $u(t_2)$. Assuming a pinhole model for the camera, then $u(t_1), u(t_2) \in \mathbb{R}^2$. Let $^cq(t_1)$ and $^cq(t_2)$ be the homogeneous representations of these locations. As standard, one can think of these vectors as the vectors from the optical-center of the camera to the projection point on the image plane. Using an initial-guess of the pose of the camera at $t_1$, the line passing through $p_E(t_1)$ and $^cq(t_1)$ can be intersected with the DTM. Any ray-tracing style algorithm can be used for this purpose. The location of this intersection is denoted as $^WG_E$. The subscript letter ``$E$'' highlights the fact that this ground-point is the estimated location for the feature point, that in general will be different from the true ground-feature location $^WG$. The difference between the true and estimated locations is due to two main sources: the error in the initial guess for the pose and the errors in the determination of $^WG_E$ caused by DTM discretization and intrinsic errors. For a reasonable initial-guess and DTM-related errors, the two points $^WG_E$ and $^WG$ will be close enough so as to allow the linearization of the DTM around $^WG_E$. Denoting by $N$ the normal of the plane tangent to the DTM at the point $^WG_E$, one can write:

$$N^T(^WG - {^WG_E}) \approx 0 \qquad (1)$$

The true ground feature $^WG$ can be described using true pose parameters:

$$^WG = R(t_1) \cdot {^cq(t_1)} \cdot \lambda + p(t_1) \qquad (2)$$

Here, $\lambda$ denotes the depth of the feature point (i.e. the distance of the point to the image plane projected on the optical-axis). Replacing (2) in (1):

$$N^T(\lambda \cdot R(t_1) \cdot {^cq(t_1)} + p(t_1) - {^WG_E}) = 0 \qquad (3)$$

From this expression, the depth of the true feature can be computed using the estimated feature location:

$$\lambda = \frac{N^{T\,W}G_E - N^T p(t_1)}{N^T R(t_1) {^cq(t_1)}} \qquad (4)$$

By plugging (4) back into (2) one gets:

$$^WG = R(t_1) {^cq(t_1)} \cdot \left( \frac{N^{T\,W}G_E - N^T p(t_1)}{N^T R(t_1) {^cq(t_1)}} \right) + p(t_1) \qquad (5)$$

In order to simplify notations, $R(t_i)$ will be replaced by $R_i$ and likewise for $p(t_i)$ and $q(t_i)$ $i=1,2$. $\Delta R(t_1,t_2)$ and $\Delta p(t_1,t_2)$ will be replaced by $R_{12}$ and $p_{12}$ respectively. The superscript describing the coordinate frame in which the vector is given will also be omitted, except for the cases were special attention needs to be drawn to the frames. Normally, $p_{12}$ and $q$'s are in camera's frame while the rest of the vectors are given in the world's frame. Using the simplified notations, (5)

can be rewritten as:
$$G = \frac{R_1 q_1 N^T}{N^T R_1 q_1} G_E - \frac{R_1 q_1 N^T}{N^T R_1 q_1} p_1 + p_1 \quad (6)$$

In order to obtain simpler expressions, define the following projection operator:
$$\mathcal{P}(u,s) \doteq \left( I - \frac{u s^T}{s^T u} \right) \quad (7)$$

This operator projects a vector onto the subspace normal to $s$, along the direction of $u$. As an illustration, it is easy to verify that $s^T \cdot \mathcal{P}(u,s) v \equiv 0$ and $\mathcal{P}(u,s) u \equiv 0$. By adding and subtracting $G_E$ to (6), and after reordering:

$$G = G_E + \left[ I - \frac{R_1 q_1 N^T}{N^T R_1 q_1} \right] p_1 - \left[ I - \frac{R_1 q_1 N^T}{N^T R_1 q_1} \right] G_E \quad (8)$$

Using the projection operator, (8) becomes:
$$G = G_E + \mathcal{P}(R_1 q_1, N)(p_1 - G_E) \quad (9)$$

The above expression has a clear geometric interpretation (see Fig.1). The vector from $G_E$ to $p_1$ is being projected onto the tangent plane. The projection is along the direction $R_1 q_1$, which is the direction of the ray from the camera's optical-center ($p_1$), passing through the image feature.

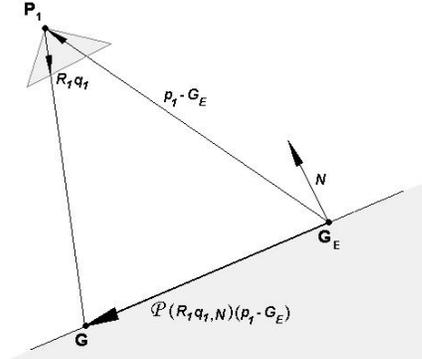

Figure 1: Geometrical description of expression (9) using the projection operator (7)

Our next step will be transferring $G$ from the global coordinates frame- $W$ into the first camera's frame $C_1$ and then to the second camera's frame $C_2$. Since $p_1$ and $R_1$ describe the transformation from $C_1$ into $W$, we will use the inverse transformation:

$$^{C_2}G = p_{12} + R_{12} \left( R_1^T (G - p_1) \right) \quad (10)$$

Assigning (9) into (10) gives:
$$^{C_2}G = p_{12} + R_{12} \mathcal{L} (G_E - p_1) \quad (11)$$

$\mathcal{L}$ in the above expression represents:
$$\mathcal{L} = \frac{q_1 N^T}{N^T R_1 q_1} \quad (12)$$

One can think of $\mathcal{L}$ as an operator with inverse characteristic to $\mathcal{P}$: it projects vectors on the ray continuing $R_1 q_1$ along the plane orthogonal to $N$.

$q_2$ is the projection of the true ground-feature $G$. Thus, the vectors $q_2$ and $^{C_2}G$ should

coincide. This observation can be expressed mathematically by projecting $^{c_2}G$ on the ray continuation of $q_2$:

$$^{c_2}G = \frac{q_2}{|q_2|} \cdot \left( \frac{q_2^T}{|q_2|} \cdot {^{c_2}G} \right) \tag{13}$$

In expression (13), $q_2^T/|q_2| \cdot {^{c_2}G}$ is the magnitude of $^{c_2}G$'s projection on $q_2$. By reorganizing (13) and using the projection operator, we obtain:

$$\left[ I - \frac{q_2 \cdot q_2^T}{q_2^T \cdot q_2} \right] \cdot {^{c_2}G} \ = \ \mathcal{P}(q_2, q_2) \cdot {^{c_2}G} \ = \ 0 \tag{14}$$

$^{c_2}G$ is being projected on the orthogonal complement of $q_2$. Since $^{c_2}G$ and $q_2$ should coincide, this projection should yield the zero-vector. Plugging (11) into (14) yields our final constraint:

$$\mathcal{P}(q_2, q_2)[p_{12} + R_{12}\mathcal{L}(G_E - p_1)] = 0 \tag{15}$$

This constraint involves the position, orientation and the ego-motion defining the two frames of the camera. Although it involves 3D vectors, it is clear that its rank can not exceed two due to the usage of $\mathcal{P}$ which projects $\mathbb{R}^3$ on a two-dimensional subspace.

Such constraint can be established for each vector in the optical-flow field, until a non-singular system is obtained. Since twelve parameters need to be estimated (six for pose and six for the ego-motion), at least six optical-flow vectors are required for the system solution. But it is correct conclusion for nonlinear problem. If we use Gauss-Newton iterations method and so make linearization of our problem near approximate solution. The found matrix will be always singular for six points (with zero determinant)as numerical simulations demonstrate. So it is necessary to use at least seven points to obtain nonsingular linear approximation. Usually, more vectors will be used in order to define an over-determined system, which will lead to more robust solution. The reader attention is drawn to the fact that a non-linear constraint was obtained. Thus, an iterative scheme will be used in order to solve this system. A robust algorithm which uses Gauss-Newton iterations and M-estimator is described in [10].We begin to use Levenberg-Marquardt method if Gauss-Newton method after several iterations stopped to converge. This two algorithms are realized in lsqnonlin() Matlab function. The applicability, accuracy and robustness of the algorithm was verified though simulations and lab-experiments.

It is more convenient to use more robust for iterations equivalent to (15) equation:

$$\mathcal{P}(q_2, q_2)[p_{12} + R_{12}\mathcal{L}_i(G_{E_i} - p_1)] / |^{c_2}G| = 0 \tag{16}$$

Using of this normalized form of equations avoids to get incorrect trivial solution when two positions are in a single point on the ground.

### 3.1 Multiple Features

Suppose next that $n$ feature points are tracked in two frames, so that the estimated locations $Q_{Ei}$ and projections onto the image plane $q_{1i}$ and $q_{2i}$ are estimated and measured, respectively, for $i = 1, \cdots, n$. Associated with each $Q_{Ei}$ is the normal vector to the DTM at this point, namely $N_i$.

Taking this into account, one can re-write (15) in matrix form as:

$$\left[-\mathcal{P}(q_{2i}) \quad \mathcal{P}(q_{2i})\frac{R_{12}q_{1i}N_i^T}{N_i^T R_1 q_{1i}}\right]\begin{bmatrix}p_{12}\\p_1\end{bmatrix} =$$

$$\mathcal{P}(q_{2i})\frac{R_{12}q_{1i}N_i^T}{N_i^T R_1 q_{1i}}Q_{Ei}. \tag{17}$$

Repeating this for each feature point:

$$\begin{bmatrix}-\mathcal{P}(q_{21}) & \mathcal{P}(q_{21})\frac{R_{12}q_{11}N_1^T}{N_1^T R_1 q_{11}}\\ -\mathcal{P}(q_{22}) & \mathcal{P}(q_{22})\frac{R_{12}q_{12}N_2^T}{N_2^T R_1 q_{12}}\\ \vdots & \vdots \\ -\mathcal{P}(q_{2n}) & \mathcal{P}(q_{2n})\frac{R_{12}q_{1n}N_n^T}{N_n^T R_1 q_{1n}}\end{bmatrix}\begin{bmatrix}p_{12}\\p_1\end{bmatrix} =$$

$$\begin{bmatrix}\mathcal{P}(q_{21})\frac{R_{12}q_{11}N_1^T}{N_1^T R_1 q_{11}}Q_{E1}\\ \mathcal{P}(q_{22})\frac{R_{12}q_{12}N_2^T}{N_2^T R_1 q_{12}}Q_{E2}\\ \vdots \\ \mathcal{P}(q_{2n})\frac{R_{12}q_{1n}N_n^T}{N_n^T R_1 q_{1n}}Q_{En}\end{bmatrix} \tag{18}$$

In compact notation:

$$A_n \begin{bmatrix}p_{12}\\p_1\end{bmatrix} = B_n. \tag{19}$$

Note that $A_n$ and $B_n$ depend on known quantities: the estimated features, the normals of the DTM tangent planes, and the images of the features at the two time instances, together with the unknown orientation $R_1$ and the relative rotation $R_{12}$. At this point in our discussion, several remarks are in order.

<u>Remark 1</u>: The constraint (18) involves twelve "unknowns", namely the pose and ego-motion of the camera. From the remark at the end of the previous section, the equation involves at most $2n$ linearly independent constraints, so that at least six features at different locations $Q_{Ti}$ are required to have a determinate system of equations. But it is correct conclusion for nonlinear problem. If we use Gauss-Newton iterations method and so make linearization of our problem near approximate solution. The found matrix will be always singular for six points (with zero determinant)as numerical simulations demonstrate. So it is necessary to use at least seven points to obtain nonsingular linear approximation. Usually, more vectors will be used in order to define an over-determined system, and hence reduce the effect of noise. Clearly, there are degenerate scenarios in which the obtained system is singular, no matter what is the number of available features. Examples for such scenarios include flying above completely planar or spherical terrain. However, in the general case where the terrain has ``interesting'' structure the system is non-singular and the twelve parameters can be obtained.

<u>Remark 2</u>: The constraint (18) is non-linear and, therefore, no analytic solution to it is readily available. Thus, an iterative scheme will be used in order to solve this system. A robust algorithm using Newton-iterations and M-estimator will be described in following sections.

<u>Remark 3</u>: Given Remark 2, one observes that the location and translation appear linearly in the constraint. Using the pseudo-inverse, these two vectors can be solved explicitly to give:

$$\begin{bmatrix} p_{12} \\ p_1 \end{bmatrix} = A_n^\dagger B_n, \tag{20}$$

so that, after resubstituting in (19):

$$(I - A_n A_n^\dagger) B_n = 0. \tag{21}$$

This remark leads to two conclusions:

1. If the rotation is known to good accuracy and measurement noise is relatively low, then the position and translation can be determined by solving a linear equation. This fact may be relevant when "fusing" the procedure described here with other measurement, e.g., with inertial navigation.

2. Equation (21) shows that the estimation of rotation (both absolute and relative) can be separated from that of location/translation. This fact is also found when estimating pose from a set of visible landmarks as shown in [11]. In that work, similarly to the present, the estimate is obtained by minimizing an objective function which measures the errors in the *object-space* rather than on the image plane (as in most other works). This property enables the decoupling of the estimation problem. Note however that [11] address's only the pose rotation and translation decoupling while here the 6 parameters of absolute and relative rotations are separated from the 6 parameters of the camera location and translation.

## 3.2 The Epipolar Constraint Connection

Before proceeding any further, it is interesting to look at (15) in the light of previous work in SFM and, in particular, epipolar geometry. In order to do this, it is worth deriving the basic constraint in the present framework and notation. Write:

$$^{c_2}Q_T = \lambda_2 q_2 = p_{12} + \lambda_1 R_{12} q_1 \tag{22}$$

for some scalars $\lambda_1$ and $\lambda_2$ (see Fig.2).

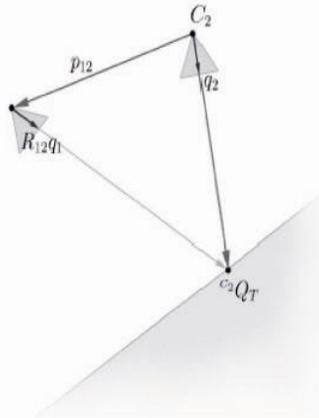

Figure 2: The examined scenario from the second camera frame's ($C_2$) point of view. $q_2$ is the perspective projection of the terrain feature $^{C_2}Q_T$, and thus the two should coincide. Additionally, since $q_1$ is also a projection of the same feature in the $C_1$-frame, the epipolar constraint requires that the two rays (one in the direction of $q_2$ and the other from $p_{12}$ in the direction of $R_{12}q_1$) will intersect.

It follows that:
$$p_{12} \times \lambda_2 q_2 = p_{12} \times \lambda_1 R_{12} q_1, \tag{23}$$
and hence:
$$q_2^T(p_{12} \times R_{12} q_1) = 0. \tag{24}$$
For a vector $x \in \mathbb{R}^3$, let $x^\wedge$ denote the skew-symmetric matrix:
$$x^\wedge = \begin{bmatrix} x_1 \\ x_2 \\ x_3 \end{bmatrix}^\wedge = \begin{bmatrix} 0 & -x_3 & x_2 \\ x_3 & 0 & -x_1 \\ -x_2 & x_1 & 0 \end{bmatrix}$$
Then, it is well known that the vector product between two vectors $x$ and $y$ can be expressed as:
$$x \times y = x^\wedge y.$$
Using this notation, the epipolar constraint (24) can be written as:
$$q_2^T (R_{12} q_1)^\wedge p_{12} = 0 \tag{25}$$
and symmetrically as:
$$q_1^T R_{12}^T q_2^\wedge p_{12} = 0 \tag{26}$$
The important observation here is that if the vector $p_{12}$ verifies the above constraint, then the vector $\kappa \cdot p_{12}$ also verifies the constraint, for any number $\kappa$. This is an expression of the ambiguity built into the SFM problem. On the other hand, the constraint (15) is non-homogeneous and hence does not suffer from the same ambiguity. In terms of the translation alone (and for only one feature point!), if $p_{12}$ verifies (15) for given $R_1$ and $R_{12}$, then also $p_{12} + \kappa q_2$ will verify the constraint, and hence the ego-motion translation is defined up to a one-dimensional vector. However, one has the following trivially:
$$q_1^T R_{12}^T q_2^\wedge q_2 = 0, \tag{27}$$
and hence the epipolar constraint does not provide an additional equation that would allow us to solve for the translation in a unique manner. Moreover, observe that (15) can be written using a vector product instead of the projection operator as:
$$q_2^\wedge \left[ p_{12} + \frac{R_{12} q_1 N^T}{N^T R_1 q_1} (Q_E - p_1) \right] = 0. \tag{28}$$

Taking into account the identity
$$(R_{12} q_1)^T q_2^\wedge R_{12} q_1 \equiv 0, \tag{29}$$
it is possible to conclude that (28) → (26), and hence the new constraint "contains" the classical epipolar geometry. Indeed, one could think of the constraint derived in (15) as strengthening the epipolar constraint by requiring not only that the two rays (in the directions of $q_1$ and $q_2$) should intersect, but, in addition, that this intersection point should lie on the DTM's linearization plane. Observe, moreover, that taking more than one feature point would allow us to completely compute the translation (at least for the given rotation matrices).

## 4. Vision-based navigation algorithm corrections for inertial navigation by help of Kalman filter

Vision-based navigation algorithms has been a major research issue during the past decades. Algorithm used in this paper is based on foundations of multiple-view geometry and a land map. By help of this method we get position and orientation of a observer camera. On the other hand we obtain the same data from inertial navigation methods. To adjust these two results Kalman filter is used. We employ in this paper extended Kalman filter for nonlinear equations [12].

For inertial navigation computations was used Inertial Navigation System Toolbox for Matlab [13].

Input of Kalman filter consists of two part. The first one is variables $X$ for equations of motion. In our case it is inertial navigation equations. Vector $X$ consists of fifteen components: $[\delta x\, \delta y\, \delta z\, \delta V_x\, \delta V_y\, \delta V_z\, \delta\phi\, \delta\theta\, \delta\psi\, a_x\, a_y\, a_z\, b_x\, b_y\, b_z]$. Coordinates $\delta x\, \delta y\, \delta z$ are defined by difference between real position of the camera and position gotten from inertial navigation calculus. Variables $\delta V_x\, \delta V_y\, \delta V_z$ are defined by difference between real velocity of the camera and velocity gotten from inertial navigation calculus. Variable $\delta\phi\, \delta\theta\, \delta\psi$ are defined as Euler angles of matrix $D_r * D_c^T$ where $D_r$ is matrix defined by real Euler angles of camera with respect to Local Level Frame (L-Frame) and $D_c$ is matrix defined by Euler angles of camera with respect to Local Level Frame (L-Frame) gotten by inertial navigation computation. It is necessary to pay attention that found Euler angles $\delta\phi\, \delta\theta\, \delta\psi$ ARE NOT equivalent to difference between real Euler angles and Euler angles gotten from inertial navigation calculus. For small values of $\delta\phi\, \delta\theta\, \delta\psi$ perturbations to these angles can be added linearly and so these angles can be used in Kalman filter for small errors. Such choose of angles is made because formulas describing their evolution are much simpler than formulas describing evolution of Euler angles differences. Variables $a_x\, a_y\, a_z$ are defined by vector of Accel bias in inertial navigation measurements. Variables $b_x\, b_y\, b_z$ are defined by vector of Gyro bias in inertial navigation measurements.

The second input of Kalman filter is $Z$-result of measurements by vision-based navigation algorithms. Vector $Z$ consists of six components $[\delta x_m\, \delta y_m\, \delta z_m\, \delta\phi_m\, \delta\theta_m\, \delta\psi_m]$ Coordinates $\delta x_m\, \delta y_m\, \delta z_m$ are difference between camera position measured by vision-based navigation algorithm and position gotten from inertial navigation calculus. Variable $\delta\theta_m\, \delta\psi_m$ are defined as Euler angles of matrix $D_m * D_c^T$ where $D_m$ is matrix defined by Euler angles of camera with respect to Local Level Frame (L-Frame) measured by vision-based navigation algorithm and $D_c$ is matrix defined by Euler angles of camera with respect to Local Level Frame (L-Frame) gotten by inertial navigation computation. Let variable $k$ to be number of step for time discretization used in Kalman filter.

We assume that errors for between values gotten by inertial navigation computation and real values are linearly depend on noise. Corespondent process noise covariance matrix is denoted by $Q_k$. Diagonal elements of $Q_k$ correspondent to velocity are defined by Accel noise and proportional to $dt^2$: $Q_V \sim dt^2$, where $dt$ is time interval between $t_k$ and $t_{k-1}$: $dt = t_k - t_{k-1}$. Diagonal elements of $Q_k$ correspondent to Euler angles are defined by Gyro noise and proportional to $dt$: $Q_A \sim dt$.

We assume that errors for between values gotten by vision-based navigation algorithm and real values are linearly depend on noise. Corespondent measurement noise covariance matrix is denoted by $R_k$. Error analysis giving this matrix is described in [14].

Kalman filter equations describe evolution of *a posteriori* state estimation $X_k$ described above and *a posteriori* error covariation covariance matrix $P_k$ for variables $X_k$.

To write Kalman filter equations we must define two 15x15 matrices yet: $H_k$ and $A_k$. Matrix $H_k$ is measurement Jacobian describing connection between predicted measurement $H_k * X_k$ and actual measurement $Z_k$ defined above. Diagonal elements $H_k(1,1)$, $H_k(2,2)$, $H_k(3,3)$ describing coordinate and elements $H_k(4,7)$, $H_k(5,8)$, $H_k(6,9)$ describing angles are equal to one. The rest of the elements are equal to zero.

$A_k$ is Jacobian matrix describing evolution of vector $X_k$. The exact expression for this matrix is very difficult so we use approximate formula for $A_k$ neglecting by Coriolis effects, Earth rotation and so on. Let $\phi\theta\psi$ be the Euler angles in L-Frame, $dV$ is deltaV vector gotten from inertial navigation measurements, $f_{vec}$ is acceleration vector in L-frame, $DCM_{b-to-l}$ is direction cosine matrix (from body-frame to L-frame).

The formulas defining $A_k$ are follow:

$$\Psi_{DCM} = \begin{pmatrix} cos(\psi) & sin(\psi) & 0 \\ -sin(\psi) & cos(\psi) & 0 \\ 0 & 0 & 1 \end{pmatrix} \quad (30)$$

$$\Theta_{DCM} = \begin{pmatrix} cos(\theta) & 0 & -sin(\theta) \\ 0 & 1 & 0 \\ sin(\theta) & 0 & cos(\theta) \end{pmatrix} \quad (31)$$

$$\Phi_{DCM} = \begin{pmatrix} 1 & 0 & 0 \\ 0 & cos(\phi) & sin(\phi) \\ 0 & -sin(\phi) & cos(\phi) \end{pmatrix} \quad (32)$$

$$DCM_{b-to-l} = \Phi_{DCM}\Theta_{DCM}\Psi_{DCM} \quad (33)$$

$$f_{vec} = DCM_{b-to-l}\frac{dV}{dt} \quad (34)$$

$$Phi(1:3,4:6) = \begin{pmatrix} 1 & 0 & 0 \\ 0 & 1 & 0 \\ 0 & 0 & 1 \end{pmatrix} \quad (35)$$

$$Phi(4:6,7:9) = \begin{pmatrix} 0 & -f_{vec}(3) & f_{vec}(2) \\ f_{vec}(3) & 0 & -f_{vec}(1) \\ -f_{vec}(2) & f_{vec}(1) & 0 \end{pmatrix} \qquad (36)$$

$$Phi(7:9,10:12) = -DCM_{b-to-l} \qquad (37)$$

$$Phi(4:6,13:15) = -DCM_{b-to-l} \qquad (38)$$

The rest of elements for matrix Phi are equal to zero.

$$A_k = I + Phi\, dt \qquad (39)$$

Kalman filter time update equations are follow:

$$X_k^- = [0\,0\,0\,0\,0\,0\,0\,0\,0\, a_{xk-1}\, a_{yk-1}\, a_{zk-1}\, b_{xk-1}\, b_{yk-1}\, b_{zk-1}] \qquad (40)$$

$$P_k^- = A_k P_{k-1} A_k^T + Q_{k-1} \qquad (41)$$

Kalman filter update equations project the state and covariance estimates from the previous time step $k-1$ to the current time step $k$.

Kalman filter measurement update equations are follow:

$$K_k = P_k^- H_k^T (H_k P_k^- H_k^T + R_k)^{-1} \qquad (42)$$

$$X_k = X_k^- + K_k (Z_k - H_k X_k^-) \qquad (43)$$

$$P_k = (I - K_k H_k) P_k^- (I - K_k H_k)^T + K_k R_k K_k^T \qquad (44)$$

Kalman filter measurement update equations correct the state and covariance estimates with measurement $Z_k$.

The found vector $X_k$ is used to update coordinates, velocities, Euler angles, Accel and Gyro biases for inertial navigation calculations on the next step.

Numerical simulations were realized to examine effectiveness of Kalman filter to combine these two navigation algorithms. On Figs. 3-5 we can see that corrected path for coordinate error much smaller than inertial navigation coordinate error without Kalman filter. Improved results by help Kalman filter are gotten also for velocity in spite of the fact that this velocity was not measured by help vision-based navigation algorithm (Fig. 6).

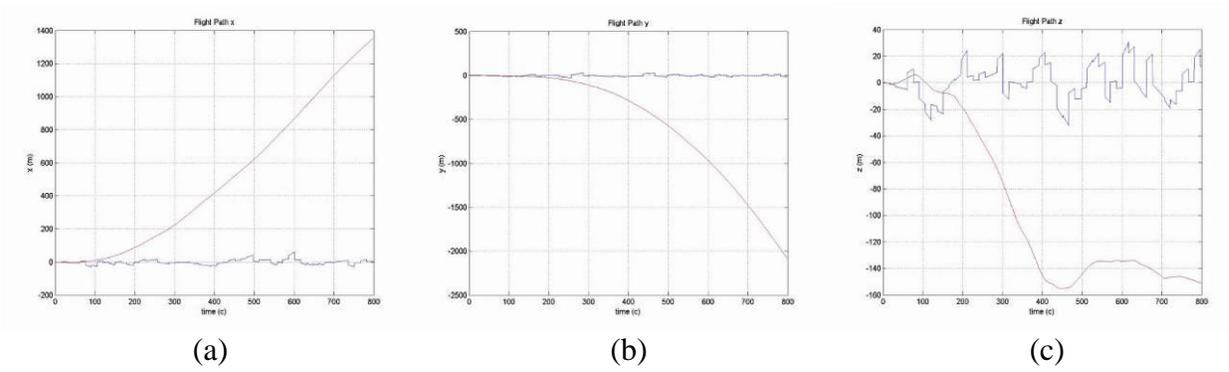

Figure 3: Position errors ((a) for x coordinate (b) for y coordinate (c) for z coordinate) of the drift path are marked with a red line, and errors of the corrected path are marked with a blue line. Parameters : Height 1000m, FOV 60 degree, Features number 120, Resolution 1000x1000, Baseline=200m, $\Delta time = 15$ s

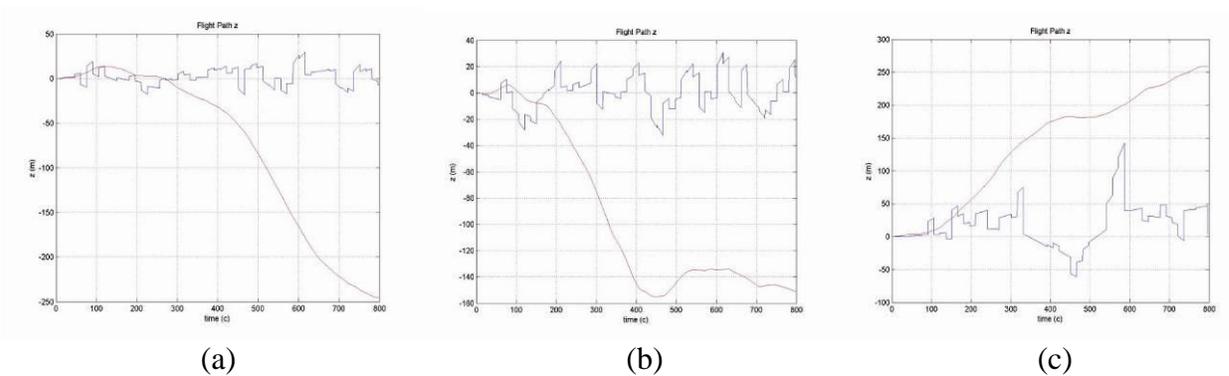

Figure 4: Position errors for z coordinate of the drift path are marked with a red line, and errors of the corrected path are marked with a blue line. Parameters : FOV 60 degree, Features number 120, Resolution 1000x1000, Baseline=200m, $\Delta time = 15$ s, Height a) 700m b) 1000m c) 3000m

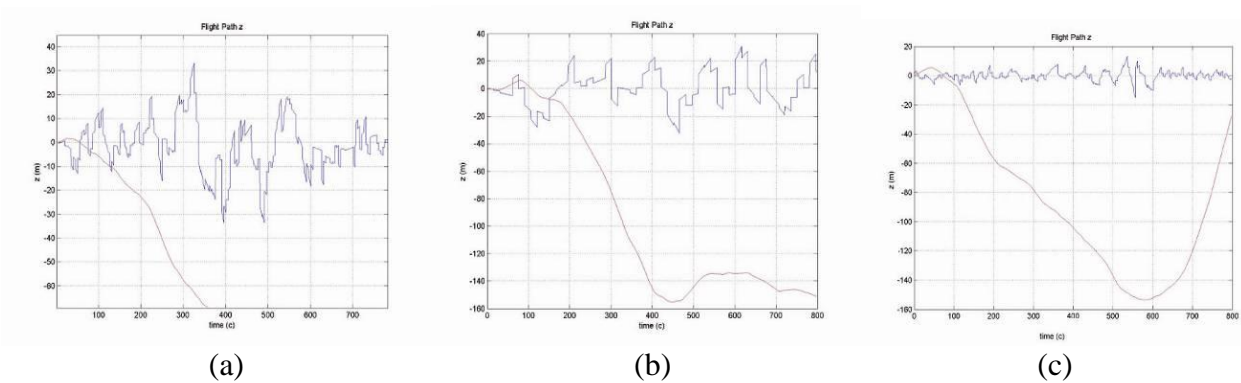

(a) (b) (c)

Figure 5: Position errors for z coordinate of the drift path are marked with a red line, and errors of the corrected path are marked with a blue line. Parameters : FOV 60 degree, Features number 120, Baseline=200m, $\Delta time = 15$ s, Height 1000m, Resolution a) 500x500 b) 1000x1000 c) 4000x4000

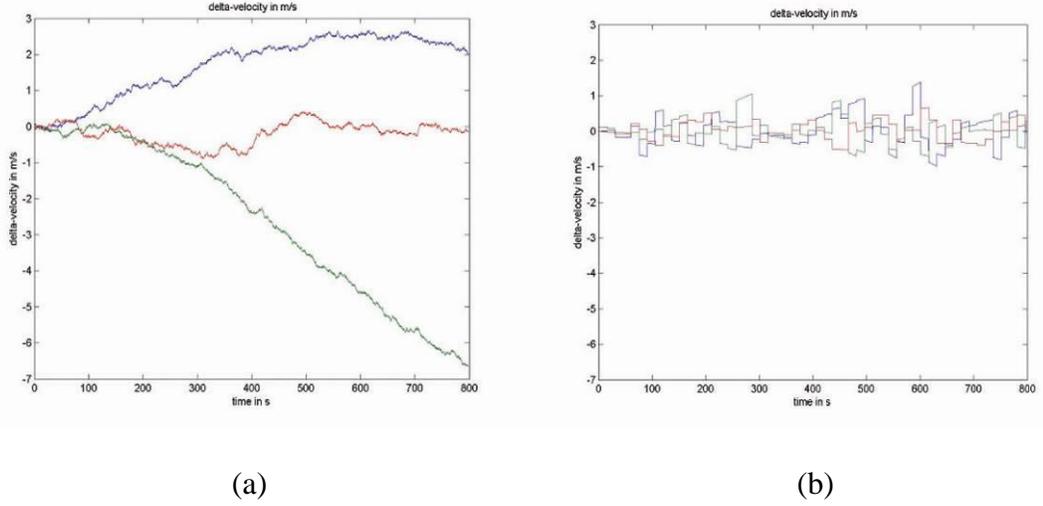

(a)  (b)

Figure 6: (a) Velocity errors of the drift path (x y z components), and (b) Velocity errors of the corrected path (x y z components). Parameters  Height 1000m, FOV 60 degree, Features number 120, Resolution 1000x1000, Baseline=200m, $\Delta time = 15$ s

## 5. Error analysis

The rest of this work deals with the error-analysis of the proposed algorithm. In order to evaluate the algorithm's performance, the objective-function of the minimization process needs to be defined first: For each of the $n$ optical-flow vectors, the function $f_i : \mathbb{R}^{12} \rightarrow \mathbb{R}^3$ is defined as the left-hand side of the constraint described in (16):

$$f_i(p_1, \phi_1, \theta_1, \psi_1, p_{12}, \phi_{12}, \theta_{12}, \psi_{12}) =$$

$$= \mathcal{P}(q_2, q_2) \left[ p_{12} + R_{12} \mathcal{L}_i \left( G_{E_i} - p_1 \right) \right] /\vert^{c_2} G \vert \qquad (45)$$

In the above expression, $R_{12}$ and $\mathcal{L}_i$ are functions of $(\phi_{12}, \theta_{12}, \psi_{12})$ and $(\phi_1, \theta_1, \psi_1)$ respectively. Additionally, the function $F : \mathbb{R}^{12} \rightarrow \mathbb{R}^{3n}$ will be defined as the concatenation of the $f_i$ functions: $F(p_1, \phi_1, \theta_1, \psi_1, p_{12}, \phi_{12}, \theta_{12}, \psi_{12}) = [f_1, \ldots, f_n]^T$. According to these notations, the goal of the algorithm is to find the twelve parameters that minimize $M(\theta, D) = \Vert F(\theta, D) \Vert^2$, where $\theta$ represents the 12-vector of the parameters to be estimated, and $D$ is the concatenation of all the data obtain from the optical-flow and the DTM. If $D$ would have been free of errors, the true parameters were obtained. Since $D$ contains some error perturbation, the estimated parameters are drifted to erroneous values. It has been shown in [15] that the connection between the uncertainty of the data and the uncertainty of the estimated parameters can be described by the following first-order approximation:

$$\Sigma_\theta = \left(\frac{dg}{d\theta}\right)^{-1} \left(\frac{dg}{dD}\right) \Sigma_D \left(\frac{dg}{dD}\right)^T \left(\frac{dg}{d\theta}\right)^{-1} \tag{46}$$

Here, $\Sigma_\theta$ and $\Sigma_D$ represent the covariance matrices of the parameters and the data respectively. $g$ is defined as follows:

$$g(\theta, D) := \frac{d}{d\theta} M(\theta, D) = \frac{d}{d\theta} F^T F = 2 J_\theta^T F \tag{47}$$

$J_\theta = dF/d\theta$ is the $(3n \times 12)$ Jacobian matrix of $F$ with respect to the twelve parameters. By ignoring second-order elements, the derivations of $g$ can be approximate by:

$$\frac{dg}{d\theta} \approx 2 J_\theta^T J_\theta \tag{48}$$

$$\frac{dg}{dD} \approx 2 J_\theta^T J_D \tag{49}$$

$J_D = dF/dD$ is defined in a similar way as the $(3n \times m)$ Jacobian matrix of $F$ with respect to the $m$ data components. Assigning (48) and (49) back into (46) yield the following expression:

$$J_T = \left(J_\theta^T J_\theta\right)^{-1} J_\theta^T$$

$$\Sigma_\theta = J_T \cdot \left(J_D \Sigma_D J_D^T\right) \cdot J_T^T \tag{50}$$

The central component $J_D \Sigma_D J_D^T$ represents the uncertainties of $F$ while the pseudo-inverse matrix $\left(J_\theta^T J_\theta\right)^{-1} J_\theta^T$ transfers the uncertainties of $F$ to those of the twelve parameters. In the following subsections, $J_\theta$, $J_D$ and $\Sigma_D$ are explicitly derived.

### 5.1 $J_\theta$ Calculation

Simple derivations of $f_i$ which is presented in (45), yield the following results:

$$N_P(q_2, {}^{c_2}G) = \mathcal{P}(q_2, q_2) \mathcal{P}({}^{c_2}G, {}^{c_2}G) / |{}^{c_2}G| \tag{51}$$

$$\frac{df}{dp_1} = -N_P(q_2, {}^{c_2}G) R_{12} \mathcal{L} \tag{52}$$

$$\frac{df}{d\alpha_1} = -N_P(q_2, {}^{c_2}G) R_{12} \mathcal{L} \left(\frac{d}{d\alpha_1} R_1\right) \mathcal{L}(G_E - p_1) \tag{53}$$

$$\frac{df}{dp_{12}} = N_P(q_2, {}^{c_2}G) \tag{54}$$

$$\frac{df}{d\alpha_{12}} = N_P(q_2, {}^{c_2}G)\left(\frac{d}{d\alpha_{12}}R_{12}\right)\mathcal{L}(G_E - p_1) \tag{55}$$

In expressions (53) and (55): $\alpha_1 = \phi_1, \theta_1, \psi_1$ and: $\alpha_{12} = \phi_{12}, \theta_{12}, \psi_{12}$. The Jacobian $J_\theta$ is obtained by simple concatenation of the above derivations.

### 5.2 $J_D$ Calculation

Before calculating $J_D$, the data vector $D$ must be explicitly defined. Two types of data are being used by the proposed navigation algorithm: data obtained from the optical-flow field and data obtained form the DTM. Each flow vector starts at $q_1$ and ends at $q_2$. One can consider $q_1$'s location as an arbitrary choice of some ground feature projection, while $q_2$ represent the new projection of the same feature on the second frame. Thus the flow errors are realized through the $q_2$ vectors.

The DTM errors influence the $G_E$ and $N$ vectors in the constraint equation. As before, the DTM linearization assumption will be used. For simplicity the derived orientation of the terrain's local linearization, as expressed by the normal, will be considered as correct while the height of this plane might be erroneous. The connection between the height error and the error of $G_E$ will be derived in the next subsection. Resulting from the above, the $q_1$'s and the $N$'s can be omitted from the data vector $D$. It will be defined as the concatenation of all the $q_2$'s followed by concatenation of the $G_E$'s.

The i'th feature's data vectors: $q_{2_i}$ and $G_{E_i}$ appears only in the i'th feature constraint, thus the obtained Jacobian matrix $J_D = [J_q, J_G]$ is a concatenation of two block diagonal matrices: $J_q$ followed by $J_G$. The i'th diagonal block element is the $3\times 3$ matrix $df_i/dq_{2_i}$ and $df_i/dG_{E_i}$ for $J_q$ and $J_G$ respectively:

$$\frac{df}{dq_2} =$$

$$\frac{-1}{\|q_2\|^2}\left[(q_2^T \cdot {}^{c_2}G)I + q_2 \cdot {}^{c_2}G^T\right]\mathcal{P}(q_2, q_2)/|{}^{c_2}G| \tag{56}$$

$$\frac{df}{dG_E} = N_P(q_2, {}^{c_2}G)R_{12}\mathcal{L} \tag{57}$$

${}^{c_2}G$ in expression (56) is the ground feature $G$ under the second camera frame as defined in (11).

### 5.3 $\Sigma_D$ Calculation

As mention above, the data-vector D is constructed from concatenation of all the $q_2$'s followed by concatenation of the $G_E$'s. Thus $\Sigma_D$ should represent the uncertainty of these elements. Since the $q_2$'s and the $G_E$'s are obtained from two different and uncorrelated processed the covariance relating them will be zero, which leads to a two block diagonal matrix:

$$\Sigma_D = \begin{bmatrix} \Sigma_q & 0 \\ 0 & \Sigma_G \end{bmatrix} \qquad (58)$$

In this work the errors of image locations and DTM height are assumed to be additive zero-mean Gaussian distributed with standard-deviation of $\sigma_I$ and $\sigma_h$ respectively. Each $q_2$ vector is a projection on the image plane where a unit focal-length is assumes. Hence, there is no uncertainty about its $z$-component. Since a normal isotropic distribution was assumed for the sake of simplicity, the covariance matrix of the image measurements is defined to be:

$$\Sigma_{q_i} = \sigma_I^2 \cdot \begin{bmatrix} 1 & & \\ & 1 & \\ & & 0 \end{bmatrix} \qquad (59)$$

and $\Sigma_q$ is the matrix with the $\Sigma_{q_i}$'s along its diagonal.

In [16] the accuracy of location's height obtained by interpolation of the neighboring DTM grid points is studied. The dependence between this accuracy and the specific required location, for which height is being interpolated, was found to be negligible. Here, the above finding was adopted and a constant standard-deviation was set to all DTM heights measurements. Although there is a dependence between close $G_E$'s uncertainties, this dependence will be ignored in the following derivations for the sake of simplicity. Thus, a block diagonal matrix is obtained for $\Sigma_G$ containing the 3×3 covariance matrices $\Sigma_{G_i}$ along its diagonal which will be derived as follows: consider the ray sent from $p_1$ along the direction of $R_1 q_1$. This ray should have intersected the terrain at $G_E = p_1 + \lambda R_1 q_1$ for some $\lambda$, but due to the DTM height error the point $\tilde{G}_E = (\tilde{x}, \tilde{y}, \tilde{h})^T$ was obtained. Let $h$ be the true height of the terrain above $(\tilde{x}, \tilde{y})$ and $H = (\tilde{x}, \tilde{y}, h)$ be the 3D point on the terrain above that location.

Using that H belongs to the true terrain plane one obtains:

$$N^T(G_E - H) = N^T(p_1 + \lambda R_1 q_1 - H) = 0 \qquad (60)$$

Extracting $\lambda$ from (60) and assigning it back to $G_E$'s expression yields:

$$G_E = p_1 + R_1 \mathcal{L}(H - p_1) \qquad (61)$$

For $G_E$'s uncertainty calculation the derivative of $G_E$ with respect to $h$ should be found:

$$\frac{dG_E}{dh} = R_1 \mathcal{L} \cdot (0 \ 0 \ 1)^T = \frac{R_1 q_1}{N^T R_1 q_1} \qquad (62)$$

The above result was obtained using the fact that the $z$-component of $N$ is 1: $N = (-\nabla DTM \ 1)^T$. Finally, the uncertainty of $G_E$ is expressed by the following covariance-matrix:

$$\Sigma_{G_i} = \left(\frac{dG_E}{dh}\right) \cdot \sigma_h^2 \cdot \left(\frac{dG_E}{dh}\right)^T = \sigma_h^2 \cdot \frac{R_1 q_1 q_1^T R_1^T}{(N^T R_1 q_1)^2} \qquad (63)$$

### 5.4 $\Sigma_{C_2}$ Calculation

The algorithm presented in this work estimates the pose of the first camera frame and the ego-motion. Usually, the most interesting parameters for navigation purpose will be the second camera frame since it reflect the most updated information about the platform location. The second pose can be obtained in a straightforward manner as the composition of the first frame pose together with the camera ego-motion:

$$p_2 = p_1 - R_1 R_{12}^T p_{12} \tag{64}$$

$$R_2 = R_1 R_{12}^T \tag{65}$$

The uncertainty of the second pose estimates will be described by a $6 \times 6$ covariance matrix that can be derived from the already obtained $12 \times 12$ covariance matrix $\Sigma_\theta$ by multiplication from both sides with $J_{C_2}$. The last notation is the Jacobian of the six $C_2$ parameters with respect to the twelve parameters mentioned above. For this purpose, the three Euler angles $\phi_2$, $\theta_2$ and $\psi_2$ need to be extracted from (65) using the following equations:

$$\phi_2 = arctan\left(\frac{R_2(2,3)}{R_2(3,3)}\right) \tag{66}$$

$$\theta_2 = arcsin(-R_2(1,3)) \tag{67}$$

$$\psi_2 = arctan\left(\frac{R_2(1,2)}{R_2(1,1)}\right) \tag{68}$$

Simple derivations and then concatenation of the above expressions yields the required Jacobian which is used to propagate the uncertainty from $C_1$ and the ego-motion to $C_2$. The found covariance matrix $\Sigma_{C_2}$ is the same as measurement covariant matrix $R_k$ described in section about Kalman filter.

$$R_k = \Sigma_{C_2} \tag{69}$$

### 6. Divergence of the method. Necessary thresholds for the method convergence

In previous Section we considered Error analysis for video navigation method. But its consideration is correct only if found solution is close to true one. If it is not true nonlinear effects can appear or even we can found incorrect local minimum. In this case the method can begins to diverge. We can obtain the such result:
1) if large number of outliers features appears.
2) if the case is close to degenerated one. In this case the position or orientation errors are too large. It can happen for example for small number of features, flat ground , small field of view of camera and all that.
3) if the initial position and orientation for iterations process are too far from true values
In the follow subsections we consider some threshold conditions which allow us to avoid the such situations.
If in some case even one of these threshold conditions is not correct we don't use for this case the correction of visual navigation method and use only usual INS result.If such situation repeats three times we stope to use the visual navigation method at all and don't use it also for the last correct case. Let us discourse these three factors in details

### 6.1 Dealing with Outliers

In order to handle real data, a procedure for dealing with outliers must be included in the implementation. The objective of the present section is to describe the current implementation, which seems to work satisfactorily in practice. Three kinds of outliers should be considered:
1. Outliers present in the correspondence solution (i.e., "wrong matches").
2. Outliers caused by the terrain shape, and
3. Outliers caused by relatively large errors between the DTM and the observed terrain.

The latter two kinds of outliers are illustrated in Fig.7. The outliers caused by the terrain shape appear for terrain features located close to large depth variations. For example, consider two hills, one closer to the camera, the other farther away, and a terrain feature $Q$ located on the closer hill. The ray-tracing algorithm using the erroneous pose may ``miss'' the proximal hill and erroneously place the feature on the distal one. Needless to say, the error between the true and estimated locations is not covered by the linearization. To visualize the errors introduced by a relatively large DTM-actual terrain mismatch, suppose a building was present on the terrain when the DTM was acquired, but is no longer there when the experiment takes place. The ray-tracing algorithm will locate the feature on the building although the true terrain-feature belongs to a background that is now visible.

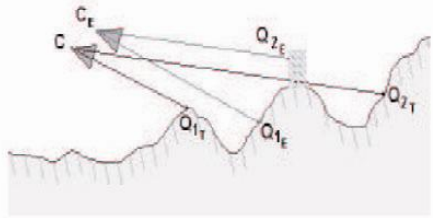

Figure 7: Outliers caused by terrain shape and DTM mismatch. $C_T$ and $C_E$ are true and estimated camera frames, respectively. $Q_{1_E}$ and $Q_{2_E}$ are outliers caused by terrain shape and by terrain/DTM mismatch, respectively.

As discussed above, the multi-feature constraint is solved in a least-squares sense for the pose and motion variables. Given the sensitivity of least-squares to incorrect data, the inclusion of one or more outliers may result in the convergence to a wrong solution. A possible way to circumvent this difficulty is by using an M-estimator, in which the original solution is replaced by a weighted version. In this version, a small weight is assigned to the constraints involving outliers, thereby minimizing their effect on the solution. More specifically, consider the function $f_i(\Theta)$ defined in (45) resulting from the $i$-th correspondence pair. In the absence of noise, this function should be equal to zero at the true pose and motion values and hence, following standard notation, define the residual $r_i(\Theta) = \|f_i(\Theta)\|$. Using an M-estimator, the solution for $\Theta$ (the twelve parameters to be estimated) is obtained using an iterative re-weighted least-squares scheme:

$$\Theta = argmin \sum_{i=1}^{n} w_i r_i^2. \quad (70)$$

The weights $w_i$ are recomputed after each iteration according to their corresponding updated residual. In our implementation we used the so-called *Geman-McClure* function, for which the weights are given by:

$$w(x) = \frac{1}{(1+x^2)^2}. \quad (71)$$

The calculated weights are then used to construct a weighted pseudo-inverse matrix that replaces the regular pseudo-inverse $J_T$ appearing in (50). See [17] for further details about M-estimation techniques. Let us define weights matrix W which allows us to decrease influence of outliers

$$r_i = \|f_i(p_1, \phi_1, \theta_1, \psi_1, p_{12}, \phi_{12}, \theta_{12}, \psi_{12})\|$$

$$medR = median(x_i)$$

$$R_i = w(r_i/medR) \quad (72)$$

where $i = 1,...,n$ and $n$ is number of features.

The weights matrix W ($3n \times 3n$) can be found as follow: for diagonal elements of W we can write : $W_{ii} = R_k$ where k is integer part of $[(i-1)/3+1]$. Non-diagonal elements of $W_{ij} = 0$ for $i \neq j$.

Instead equation (50) we use new one:

$$JT = (J_\theta^T W J_\theta)^{-1} J_\theta^T W$$

$$\Sigma_\theta = JT \cdot (J_D \Sigma_D J_D^T) \cdot JT^T \quad (73)$$

If we know two positions of camera and features position in the first photo so we can find the features position on the second photo. If the distance between true position of some correspondent feature on second photo and the position found by previously described method larger than $3\sigma_I$ we would consider the such feature as outlier. Let us define $N_i$ as number of outliers in initial approximation of cameras position and orientation (i.e. before using visual navigation method) and $N_f$ as number of outliers after visual navigation method corrections. The follow conditions let us to avoid too large number of outliers case:

$$N_i \geq N_f$$

$$\frac{N_f}{N} < threshold_\% \quad (74)$$

where N is full number of features and $threshold_\%$ is some threshold value. We choose it to be equal 0.1 .

## 6.2 Degenerated case large errors.

For degenerate case the matrix $J_\theta^T W J_\theta$ in equation (73) can be singular. It gives us follow threshold condition:

$$rcond(J_\theta^T W J_\theta) > threshold_{rcond} \qquad (75)$$

where rcond() -Matlab function for matrix reciprocal condition number estimate. It is measure for matrix singularity ($0 < rcond() < 1$). Threshold value $threshold_{rcond}$ is chosen to be $10^{-16}$.

Degenerated case because of small number of features, flat ground or small field of view of camera gives the follow threshold conditions:

$$\frac{\sqrt{[\Sigma_{C_2}]_{ii}}}{(3\sigma_I/f)h} < threshold_{dist} \qquad (76)$$

where $i = x, y, z$ coordinate indexes for diagonal elements of covariance matrix $\Sigma_{C_2}$. $f = 1$ is a focus length of the camera, h is height of the camera. $\frac{3\sigma_I}{f}h$ gives us the maximum camera position shift allowing the photo feature error to be smaller than pixel size. Threshold value $threshold_{dist}$ is chosen to be 40.

$$3\sqrt{[\Sigma_{C_2}]_{ii}} < L_{ground-dist} \qquad (77)$$

where $i = x, y, z$ coordinate indexes for diagonal elements of covariance matrix $\Sigma_{C_2}$, $L_{ground-dist}$ is character size of ground relief change.

$$\frac{\sqrt{[\Sigma_{C_2}]_{ii}}}{(3\sigma_I/f)} < threshold_{angle} \qquad (78)$$

where $i = \phi, \theta, \psi$ angular indexes for diagonal elements of covariance matrix $\Sigma_{C_2}$. $\frac{3\sigma_I}{f}$ gives us the maximum camera angular shift allowing the photo feature error to be smaller than pixel size. Threshold value $threshold_{angle}$ is chosen to be 40.

$$3\sqrt{[\Sigma_{C_2}]_{ii}} < \frac{L_{ground-dist}}{h} \qquad (79)$$

where $i = \phi, \theta, \psi$ angular indexes for diagonal elements of covariance matrix $\Sigma_{C_2}$.

Degenerated case because of small baseline (distance between two camera positions used in video navigation method) gives the follow threshold conditions:

$$\frac{\sqrt{[\Sigma_\theta]_{ii}}}{\|p_{12}\|} < threshold_{dist_{12}} \tag{80}$$

where $i = x_{12}, y_{12}, z_{12}$ mutual coordinate indexes for diagonal elements of covariance matrix $\Sigma_\theta$. Threshold value $threshold_{dist_{12}}$ is chosen to be 0.1 .

$$\frac{\sqrt{[\Sigma_\theta]_{ii}}}{(\|p_{12}\|/h)} < threshold_{angle_{12}} \tag{81}$$

where $i = \phi_{12}, \theta_{12}, \psi_{12}$ mutual angular indexes for diagonal elements of covariance matrix $\Sigma_\theta$. Threshold value $threshold_{angle_{12}}$ is chosen to be 0.1 .

### 6.3 The initial state of the camera is too far from the its true or final calculated state.

Let us define threshold conditions to avoid the initial state of the camera to be too far from the its true state. $P_k^-$ is covariant matrix obtained from INS and previous corrections of INS by video navigation method with help of Kalman filter and described in section about Kalman filter.

$$3\sqrt{[P_k^-]_{ii}} < L_{ground-dist} \tag{82}$$

where $i = x, y, z$ coordinate indexes for diagonal elements of covariance matrix $P_k^-$.

$$3\sqrt{[P_k^-]_{ii}} < \frac{L_{ground-dist}}{h} \tag{83}$$

where $i = \phi, \theta, \psi$ angular indexes for diagonal elements of covariance matrix $P_k^-$.

Let us define threshold conditions to avoid the initial state of the camera to be too far from the its final state. The follow four equations give us differences between initial and final state obtain as corrections of INS by video navigation method with help of Kalman filter.

$$\delta p_2 = |p_{2\,final} - p_{2\,init}| \tag{84}$$

$$\delta p_{12} = |p_{12\,final} - p_{12\,init}| \tag{85}$$

$$\delta \alpha_2 = |\alpha_{2\,final} - \alpha_{2\,init}| \, mod(2\pi) \tag{86}$$

$$\delta \alpha_{12} = |\alpha_{12\,final} - \alpha_{12\,init}| \, mod(2\pi) \tag{87}$$

$$3(\sqrt{[P_k^-]_{ii}} + \sqrt{[\Sigma_{C_2}]_{ii}}) > \delta p_{2i} \qquad (88)$$

where $i = x, y, z$ coordinate indexes for diagonal elements of covariance matrix $P_k^-$ and $\Sigma_{C_2}$

$$3(\sqrt{[P_k^-]_{ii}} + \sqrt{[\Sigma_{C_2}]_{ii}}) > \delta \alpha_{2i} \qquad (89)$$

where $i = \phi, \theta, \psi$ angular indexes for diagonal elements of covariance matrix $P_k^-$ and $\Sigma_{C_2}$

$$\frac{\delta p_{12i}}{\|p_{12}\|} < threshold_{dist_{12}} \qquad (90)$$

where $i = x_{12}, y_{12}, z_{12}$ mutual coordinate indexes.

$$\frac{\delta \alpha_{12i}}{(\|p_{12}\|/h)} < threshold_{angle_{12}} \qquad (91)$$

where $i = \phi_{12}, \theta_{12}, \psi_{12}$ mutual angular indexes.

## 7 Simulations Results

### 7.1 Dependence of error analysis on different factors.

The purpose of the following section is to study the influence of different factors on the accuracy of the proposed algorithm estimates. The closed form expression that was developed throughout the previous section is being used to determine the uncertainty of these estimates under a variety of simulated scenarios. Each tested scenario is characterized by the following parameters: the number of optical-flow features being used by the algorithm, the image resolution, the grid spacing of the DTM (also referred as the DTM resolution), the amplitude of hills/mountains on the observed terrain, and the magnitude of the ego-motion components. At each simulation, all parameters except the examined one are set according to a predefined parameters set. In this *default scenario*, a camera with $400 \times 400$ image resolution flies at altitude of 500m above the terrain. The terrain model dimensions are $3 \times 3$ km with 300m elevation differences (Fig.13(b)). A DTM of 30m grid spacing is being used to model the terrain (Fig.10(c)). The DTM resolution leads to a standard-deviation of 2.34m for the height measurements. The default-scenario also defines the number of optical-flow features to about 170, where an ego-motion of $\|p_{12}\| = 40m$ and $\|(\phi_{12}, \theta_{12}, \psi_{12})\| = 10°$ differs the two images being used for the optical-flow computation. Each of the simulations described below study the influence of different parameter. A variety of values are examined and 150 random tests are performed for each tested value. For each test the camera position and orientation were randomly selected, except the camera's height that was dictated by the scenario's parameters. Additionally, the direction of the ego-motion translation and rotation components were first chosen at random and then normalized to the require magnitude.

In Fig.8, the first simulation results are presented. In this simulation the number of optical-flow features that are used by the algorithm is varied and its influence on the obtained accuracy of $C_2$ and the ego-motion is studied. All parameters were set to their default values except for the features number. Fig.8(a) presents the standard-deviations of the second frame of the camera while the deviations of the ego-motion are shown in Fig.8(b). As expected, the accuracy improves as the number of features increases, although the improvement becomes negligible after the features' number reaches about 150.

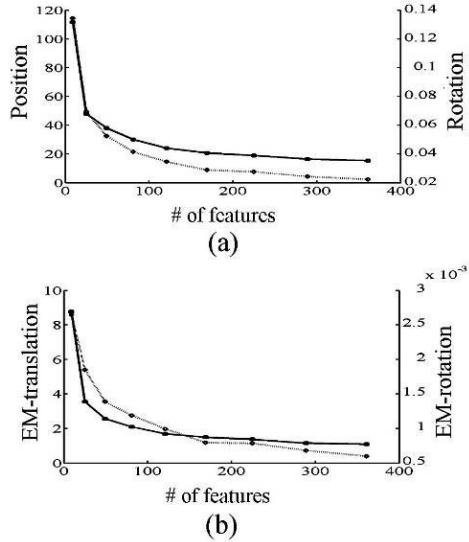

Figure 8: Average standard-deviation of the second position and orientation (a), and the ego-motion's translation and rotation (b) with respect to the number of flow-features. In both graphs, the left vertical axis measures the translational deviations (in meters) and corresponds to the solid graph-line, while the right vertical axis measures the rotational deviations (in radians) and corresponds to the dotted graph-line

In the second simulation the influence of the image resolution was studied (Fig.9). It was assumed that the image measurements contain uncertainty of half-pixel, where the size of the pixels is dictated by the image resolution. Obviously, the accuracy improves as image resolution increases since the quality of the optical-flow data is directly depends on this parameter.

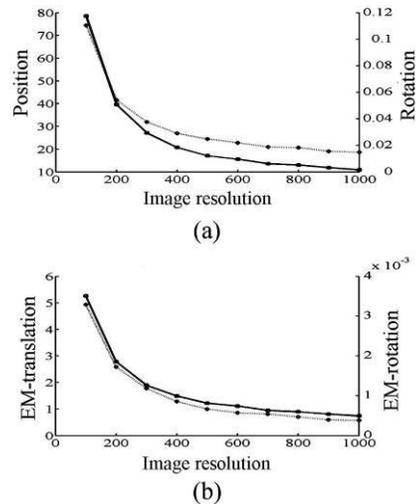

Figure 9: Average standard-deviation of the second position and orientation (a), and the ego-

motion's translation and rotation (b) with respect to the image resolution

The influence of DTM grid spacing is the objective of the next simulation. Different DTM resolutions were tested varying from 10m up to an extremely rough resolution of 190m between adjacent grid points (see Fig.10). The readers attention is drawn to the fact that the obtained accuracy seems to decrease linearly with respect to the DTM grid-spacing (see Fig.11) . This phenomenon can be understood since, as was explained in the previous section, the DTM resolution does not affect the accuracy directly but rather it influences the height uncertainty which is involved in the accuracy calculation. As can be seen in Fig.12, the standard-deviation of the DTM heights increases linearly with respect to the DTM grid spacing which is the reason for the obtained results.

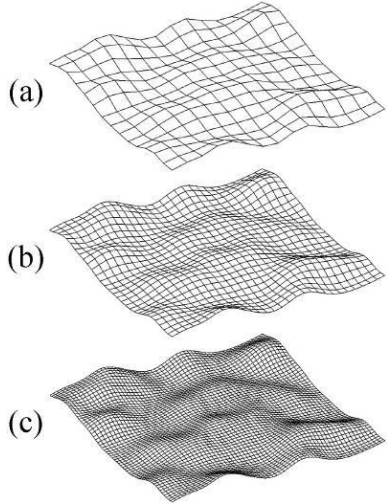

Figure 10: Different DTM resolutions: (a) grid spacing = 190m, (b) grid spacing = 100m, (c) grid spacing = 30m

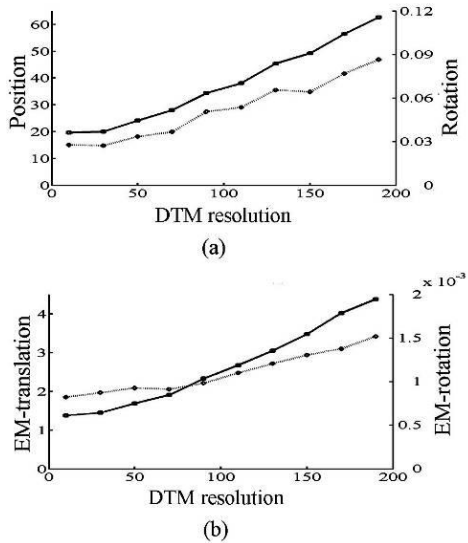

Figure 11: Average standard-deviation of the second position and orientation (a), and the ego-motion's translation and rotation (b) with respect to the grid-spacing of the DTM

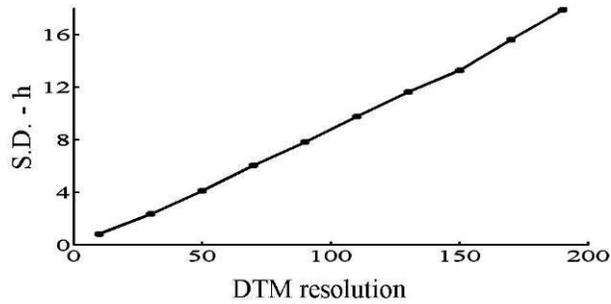

Figure 12: standard-deviation of the DTM's height measurement with respect to the grid-spacing of the DTM

Another simulation demonstrates the importance of the terrain structure to the estimates accuracy. In the extreme scenario of flying above a planar terrain, the observed ground features do not contain the required information for the camera pose derivation, and a singular system will be obtained. As the height differences and the variability of the terrain increase, the features become more informative and a better estimates can be derived. For this simulation, the DTM elevation differences were scaled to vary from 50m to 450m (Fig.13). It is emphasized that while the terrain structure plays a crucial role at the camera pose estimation together with the translational component of the ego-motion, it has no direct affect on the ego-motion rotational component. As the optical-flow is a composition of two vector fields - translation and rotation, the information for deriving the ego-motion rotation is embedded only in the rotational component of the flow-field. Since the features depths influence only the flow's translational component it is expected that the varying height differences or any other structural change in the terrain will have no affect on the ego-motion rotation estimation. The above characteristics are well demonstrated in Fig.14.

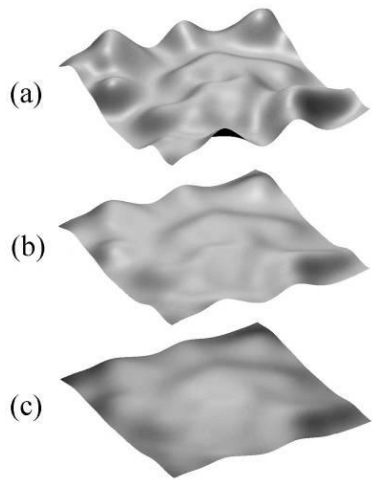

Figure 13: DTM elevation differences: (a) 150m, (b) 300m, (c) 450m

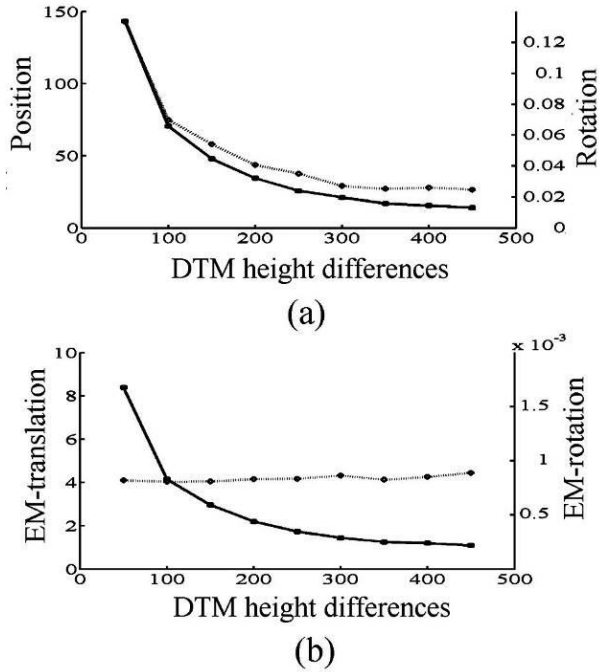

Figure 14: Average standard-deviation of the second position and orientation (a), and the ego-motion's translation and rotation (b) with respect to the height differences of the terrain

Since it is the translation component of the flow which holds the information required for the pose determination, it would be interesting to observe the effect of increasing the magnitude of this component. The last simulation presented in this work demonstrates the obtained pose accuracy when the ego-motion translation component vary form 5m to 95m. Although it has no significant effect on the ego-motion accuracy, the uncertainty of the pose estimates decreases for a large magnitude of translations (see Fig.15). As a conclusion from the above stated, the time gap between the two camera frames should be as long as the optical-flow derivation algorithm can tolerate.

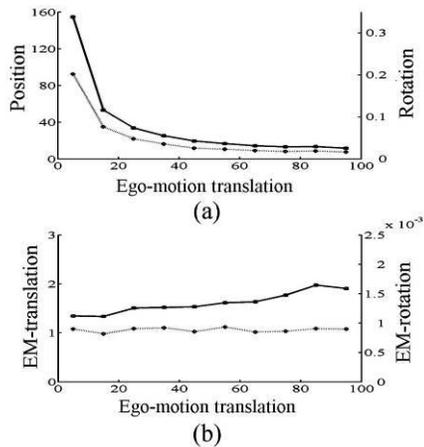

Figure 15: Average standard-deviation of the second position and orientation (a), and the ego-motion's translation and rotation (b) with respect to the magnitude of the translational component of the ego-motion

### 7.2 Results of numerical simulation for real parameters of flight and camera.

Inertial navigation systems (INS) are used usually for detection of missile position and orientation. The problem of this method is that its error increases all time. We propose to use new method (Navigation Algorithm based on Optical-Flow and a Digital Terrain Map) [18] to correct result of INS and to make the error to be finite and constant. Kalman Filter is used to combine results of INS and results of new method [12]. Error analysis with linear first-order approximation is used to find error correlation matrix for our new method [14]. We made numerical simulations of flight with real parameters of flight and camera using only INS and INS and our new method to check usefulness of this new method.

The chosen flight parameters are following:
 Height of flight is 700, 1000, 3000 m.
 Velocity of flight is 200m/s.
 Flight time is 800 s.

Trajectory of the flight we can see on (Fig.16). Digital Terrain Map of real ground was used as cell (Fig.17) for our simulations. This cell was continued periodically to obtain full Map of the ground (Fig.18). Random noise was used as main component of INS noise. The more real drift and bias noise give much bigger mistake (about 6000 m instead 1000 m in the finish point of the flight).

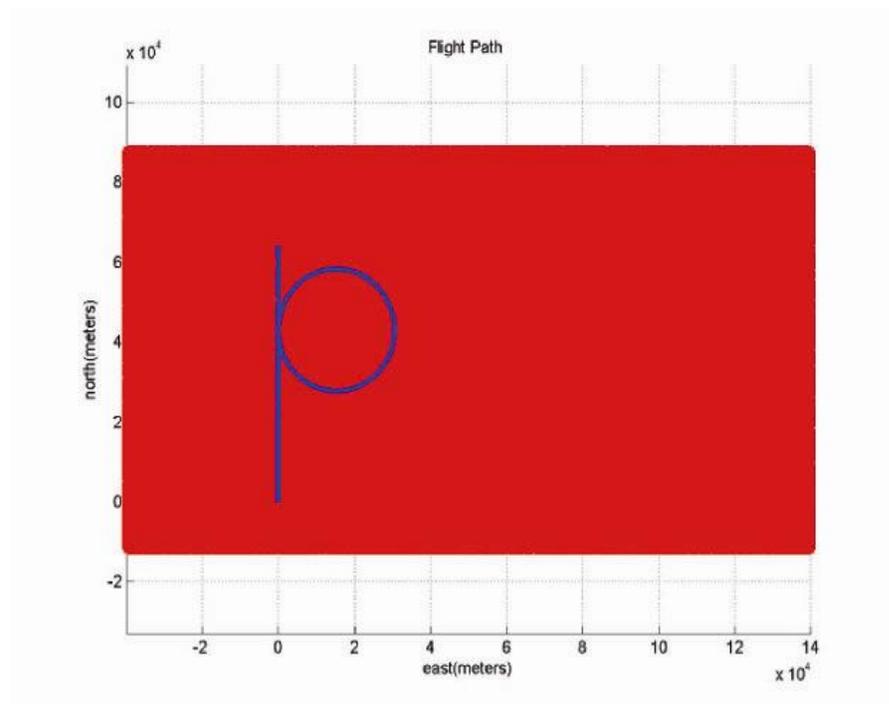

Figure 16. Trajectory of the flight.

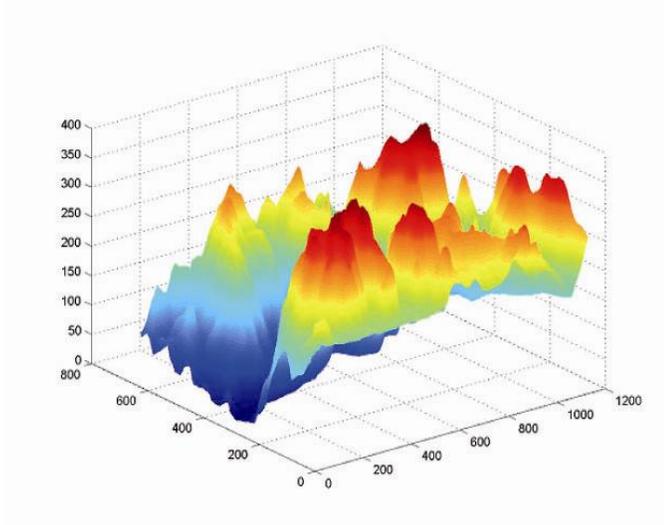

Figure 17: Map of real ground was used as cell.

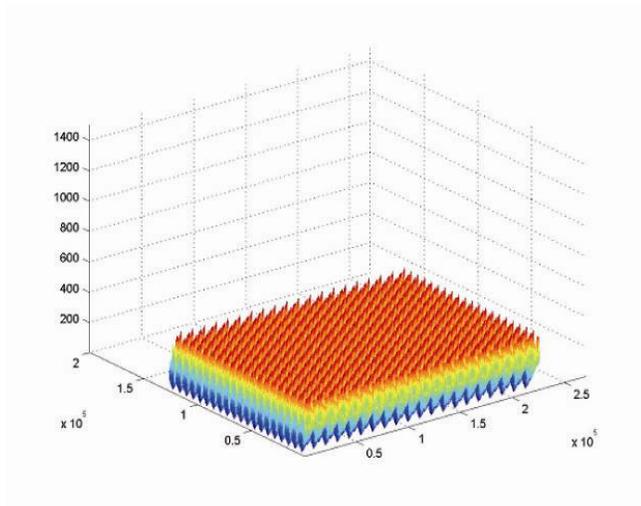

Figure 18: Cell was continued periodically to obtain full Map of the ground.

The chosen camera and simulation parameters are following:
    FOV (field of view of camera) is 60 degree. ( FOV is field of view of camera. )
    Features number found on photos is 100, 120.
    Resolution of camera is 500x500, 1000x1000, 4000x4000.( The resolution of camera defines precision of feature detection, we assume no Optical Flow outliers for features.)
    Baseline is 30m, 50m or 200m. ( Baseline is distance between two camera positions used to make two photos for new method.)
    $\Delta time$ is 5s, 15 s, 30s.( $\Delta time$ is time interval between measurements. )

The typical results of numerical simulations can be seen on (Fig.3, 4, 5, 6) for different cases of flight, camera and simulation parameters. Let us demonstrate error tables for typical case with positive results: x, y, z position errors of INS with using new method and without using new method.

Used flight, camera and simulation parameters for this case:
- FOV is 60 degree
- Number of features is 120
- Resolution is 1000x1000
- Baseline is 200m
- $\Delta time$ is 15 s.
- Flight velocity is 200 m/s
- Heights are 700m, 1000 m, 3000m.

Table 1. x axis max error for INS with and without new method for different heights.

| Height | 700m | 1000m | 3000m |
|---|---|---|---|
| Max x error without new method | 900m | 130m | 1300 m |
| Max x error with new method | 25 m | 20 m | 100 m |

Table 2. y axis max error for INS with and without new method for different heights.

| Height | 700m | 1000m | 3000m |
|---|---|---|---|
| Max y error without new method | 1000m | 2000m | 400m |
| Max y error with new method | 25m | 20m | 100 m |

Table 3. z axis max error for INS with and without new method for different heights.

| Height | 700m | 1000m | 3000m |
|---|---|---|---|
| Max z error without new method | 250m | 180m | 250 m |
| Max z error with new method | 25m | 20m | 150m |

Let us demonstrate error tables for typical case with positive results: x, y, z position errors of INS with using new method for different resolutions of camera. Used flight, camera and simulation parameters for this case:

FOV 60 degree, Number of features:120, Resolution 500x500, 1000x1000, 4000x4000, Baseline 200m, Deltatime 15 s, Flight velocity 200 m/s, Heights: 1000 m.

Table 4. x axis max error for INS with new method for different resolutions of camera.

| Resolution | 500x 500 | 1000x 1000 | 4000x 4000 |
|---|---|---|---|
| Max x error with new method | 50m | 20m | 10m |

Table 5. y axis max error for INS with new method for different resolutions of camera.

| Resolution | 500x 500 | 1000x 1000 | 4000x 4000 |
|---|---|---|---|
| Max y error with new method | 50m | 20m | 10m |

Table 6. z axis max error for INS with new method for different resolutions of camera.

| Resolution | 500x 500 | 1000x 1000 | 4000x 4000 |
|---|---|---|---|
| Max z error with new method | 35m | 20m | 10m |

## 8 Open problems and future method development.

1) If situation is close to degenerated case (for example, for small camera field of view, almost flat ground, small baseline and so on) we can not used described method because it is impossible to find cameras states from this data. But it is possible also for this case to used found correspondent features constrains for INS results improvement by help Kalman filter.We can consider directly these corespondent features (and not calculated position and orientation on basis these features) as result of measurement for Kalman filter. Example of the such improvement can be found in [19]. But in this case errors of method will increase with time similar to INS. So after some time measured position is too far from the true position and we can not use DTM constrains for error correction, but only epipolar constrains. For described in this paper method the error stops to increase and remains constant so we are capable to use DTM constrains all time.

2)It is possible to consider more optimal and fast methods for looking for minimum of function giving position and orientation of camera.For example it is possible to improve initial state for described method , using epipolar equations (25 ) for $R_{12}$ and $p_{12}$ up to constant calculations. The next step can be use equation (21) for $R_1$ calculation. And final step using equation (18) for $p_{12}$ and $p_1$ calculation.The result can be improved by described iteration method.

3)We can look for not only some random features. Also hill tops, valleys and hill occluding boundaries can be used for position and orientation specifying.

4) using distributed (not point) features and also some character object recognition.

5) Using the used methods in different practical situations: orientation in rooms, inside of man body.

## 9  Conclusions

An algorithm for pose and motion estimation using corresponding features in images and a DTM was presented with using Kalman filter. The DTM served as a global reference and its data was used for recovering the absolute position and orientation of the camera. In numerical simulations position and velocity estimates were found to be sufficiently accurate in order to bound the accumulated errors and to prevent trajectory drifts.

An error analysis has been performed for a novel algorithm that uses as input the optical flow derived from two consecutive frames and a DTM. The position, orientation and ego-motion parameters of the the camera can be estimated by the proposed algorithm. The main source for errors were identified to be the optical-flow computation, the quality of the information about the terrain, the structure of the observed terrain and the trajectory of the camera. A closed form expression for the uncertainty of the pose and motion was developed. Extensive numerical simulations were performed to study the influence of the above factors.

Tested under reasonable and common scenarios, the algorithm behaved robustly even when confronted with relatively noisy and challenging environment. Following the analysis, it is concluded that the proposed algorithm can be effectively used as part of a navigation system of autonomous vehicles.

On basis results of numerical simulation for real parameters of flight and camera we also can conclude follow:

1) The most important parameter of simulations is FOV: for the small FOV the method diverges. For FOV 60 degree the results are very good. The reason for this is that for small FOV (12 or 6 degree) the situation is close to degenerated state, also we must choose small baseline and observed ground patch is too small and almost flat.

2) Resolution of camera is also very important parameter: for better resolution we have much more better results, because of much more better precision of features detection.

3) The precision of new method depends on flight height. Initially precision increases with height increasing because we can use bigger baseline and can see bigger patch of ground. But for bigger heights precision begin to decrease because of small parallax effect.

### Acknowledgment

We would like to thank Ronen Lerner, Ehud Rivlin and Hector Rotstein for very useful consultations.